\documentclass[twocolumn]{aastex631}

\usepackage{graphicx}

\newcommand{\mum}{\ifmmode{\rm \mu m}\else{$\mu$m}\fi}

\shorttitle{Temporal spectral changes in carbon stars}
\shortauthors{Sloan et al.}

\begin{document}

\title{Temporal Changes in the Infrared Spectra of Magellanic Carbon Stars}

\correspondingauthor{G.~C. Sloan}
\email{gcsloan@stsci.edu}

\author[0000-0003-4520-1044]{G.~C.\ Sloan}
\affiliation{Space Telescope Science Institute, 3700 San Martin Drive,
             Baltimore, MD 21218, USA}
\affiliation{Department of Physics and Astronomy, University of North
             Carolina, Chapel Hill, NC 27599-3255, USA}
\author[0000-0002-2626-7155]{K.~E.\ Kraemer}
\affiliation{Institute for Scientific Research, Boston College, 140
             Commonwealth Avenue, Chestnut Hill, MA 02467, USA}
\author[0000-0001-9848-5410]{B.\ Aringer}
\affiliation{Department of Astrophysics, University of Vienna,
             T\"{u}rkenschanzstra{\ss}e 17, 1180 Wien, Austria}
\author[0000-0002-2666-9234]{J.\ Cami}
\affiliation{Department of Physics and Astronomy, The University of Western
             Ontario, London, ON N6A 3K7, Canada}
\affiliation{Institute for Earth and Space Exploration, The
             University of Western Ontario, London, ON N6A 3K7, Canada}
\author[0000-0002-1614-8195]{K.\ Eriksson}
\affiliation{Theoretical Astrophysics, Department of Physics and Astronomy,
             Uppsala University, Box 516, 751 20 Uppsala, Sweden}
\author[0000-0003-2356-643X]{S.\ H\"{o}fner}
\affiliation{Theoretical Astrophysics, Department of Physics and Astronomy,
             Uppsala University, Box 516, 751 20 Uppsala, Sweden}
\author[0000-0002-1335-5623]{E.\ Lagadec}
\affiliation{Universit\'{e} C\^{o}te d'Azur, Observatoire de la C\^{o}te
             d'Azur, CNRS, Laboratoire Lagrange, Bd de l'Observatoire,
             CS 34229, 06304 Nice Cedex 4, France}
\author[0000-0002-5529-5593]{M.\ Matsuura}
\affiliation{Cardiff Hub for Astrophysical Research and Technology (CHART),
             School of Physics and Astronomy, Cardiff University, The Parade,
             Cardiff CF24 3AA, UK}
\author[0000-0003-0356-0655]{I.\ McDonald}
\affiliation{Jodrell Bank Centre for Astrophysics, The University of
             Manchester, Manchester, M13 9PL, UK}
\author[0000-0003-2553-4474]{E.\ Montiel}
\affiliation{SOFIA-USRA, NASA Ames Research Center, MS 232-12, Moffett 
             Field, CA 94035, USA}
\author[0000-0002-6858-5063]{R.\ Sahai}
\affiliation{Jet Propulsion Laboratory, MS 183-900, California Institute of
             Technology, Pasadena, CA 91109, USA}
\author[0000-0002-3171-5469]{A.~A.\ Zijlstra}
\affiliation{Jodrell Bank Centre for Astrophysics, The University of
             Manchester, Manchester, M13 9PL, UK}

\begin{abstract}

The Medium-Resolution Spectrometer on the Mid-Infrared Instrument on 
JWST obtained spectra of three carbon stars in the Large Magellanic Cloud.  
Two of the spectra differ significantly from spectra obtained $\sim$16--19 
years earlier with the Infrared Spectrograph on the Spitzer Space 
Telescope.  The one semi-regular variable among the three has changed 
little.  The long-period Mira variable in the sample shows changes 
consistent with its pulsation cycle.  The short-period Mira shows dramatic 
changes in the strength of its molecular absorption bands, with some bands 
growing weaker and some stronger.  Whether these variations result from its 
pulsation cycle or its evolution is not clear.

\end{abstract}


\section{Introduction} 

Carbon stars form when intermediate-mass stars on the asymptotic giant 
branch (AGB) dredge up enough freshly fused carbon from their interiors 
to push the C/O ratio in their envelopes past unity \citep{hab96, hof18}.  
They dominate the dust production in the metal-poor Magellanic Clouds 
\citep{mat09, mat13, boy12}, likely making them an important contributor to 
dust in young galaxies at high redshifts.  To understand how these stars 
produce dust, it is necessary to understand the chemistry and physics of the 
carbon-bearing molecules in their envelopes and how those molecules 
condense into carbon-rich dust.

The Spitzer Space Telscope obtained infrared spectra of 144 carbon stars in
the Large Magellanic Cloud \citep[LMC; see][and references therein]{slo16}.
The majority of those spectra were observed with the low-resolution modules
on the Infrared Spectrograph \citep[IRS;][]{hou04} on the Spitzer Space
Telescope \citep{wer04}, which covered the 5--14~\mum\ region for the 
fainter targets observed, and extended to $\sim$37~\mum\ for the brighter 
targets.  The spectral resolving power ($\lambda$/$\Delta$$\lambda$) varied 
between 64 and 128, depending on wavelength, which was sufficient for 
studying solid-state emission features from dust grains and the strength and 
general profile of molecular absorption bands.

The Medium-Resolution Spectrograph \citep[MRS;][]{wel15} on the Mid-Infrared 
Instrument \citep[MIRI;][]{wri23} aboard the James Webb Space Telescope 
\citep{gar23} provides a spectral resolving power between 2000 and 3000 and 
can resolve the line structure within the absorption bands.  We obtained 
time on JWST to observe nine carbon stars in the Magellanic Clouds with the 
MRS (Program 3010).  The primary objective was to measure the resolved line 
structure to disentangle which molecules are responsible for the 
absorption at different wavelengths and to model the temperature and 
density of the absorbing gas.

This paper concentrates on the three spectra obtained prior to the
Torino XIV AGB Workshop in 2024 June.  Ultimately, this project will 
investigate the rich molecular content in the spectral data, but this 
paper focuses on the first surprise, that two of the three spectra 
obtained so far have changed significantly since they were observed with 
the IRS.

\section{The Sample} 

\begin{figure}[!ht] 
\includegraphics[width=3.4in]{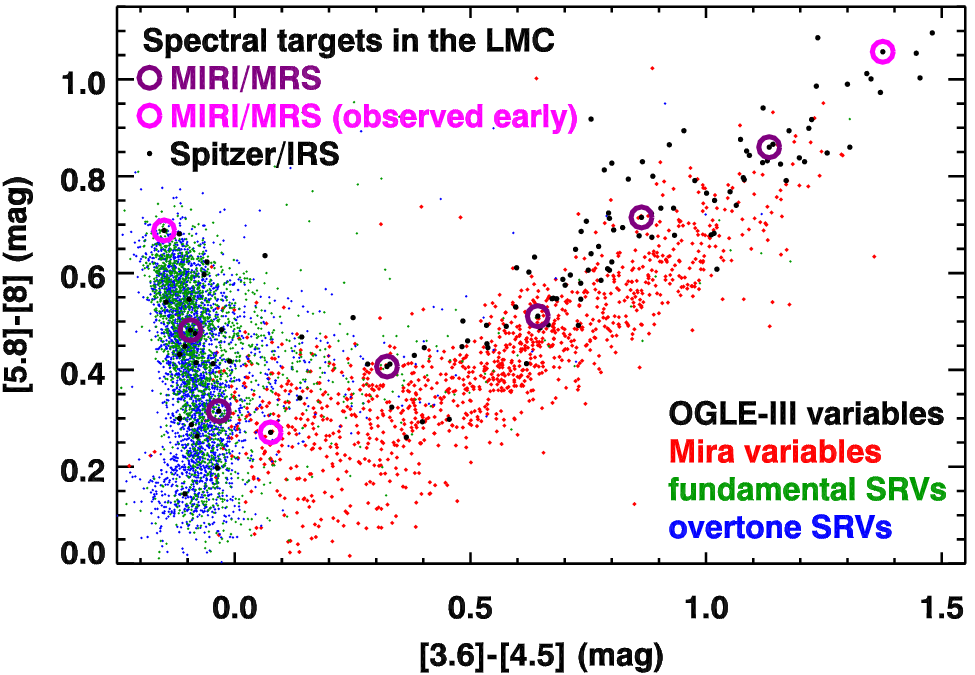}
\caption{The sample of carbon stars observed with the MRS in JWST 
program 3010 on an infrared color-color plot.  The underlying photometry is 
from WISE and Spitzer, coded by variability classes from OGLE III.  The 
three targets observed early (prior to 2024 June) are, from left to right, 
J050629, WBP~29, and MSX~LMC~736.
\label{f.cc}}
\end{figure}

Figure~\ref{f.cc} shows the full sample of nine carbon stars in our 
program and the three stars observed with the MRS between 2023 November 
and 2024 March.  Table~\ref{t.sample} provides some background information 
on the three stars observed early; they are the focus of this paper.  The 
target 2MASS J05062960$-$6855348 will be referred to as ``J050629'' 
hereafter.  The nine stars in the full sample were selected to sample two 
sequences revealed in infrared color-color space.  Figure~\ref{f.cc} 
cleanly separates the semi-regular variables (SRVs), which tend to be blue 
in most infrared colors, from the redder Miras.  The MRS sample includes 
three targets on the SRV sequence and six on the Mira sequence.

\begin{deluxetable*}{ccccccc}
\tablecolumns{7}
\tablewidth{0pt}
\tablenum{1}
\tablecaption{The first 3 carbon stars in the MRS sample}
\label{t.sample}
\tablehead{
  \colhead{Target} & \colhead{Period} & \colhead{[3.6]$-$[4.5]} &
  \colhead{[5.8]$-$[8]} & \colhead{IRS Epoch} & \colhead{MRS Epoch} &
  \colhead{$\Delta t$} \\
  \colhead{ } & \colhead{(days)} & \colhead{(mag)} & \colhead{(mag)} &
  \colhead{MJD\tablenotemark{1}} & \colhead{MJD\tablenotemark{1}} &
  \colhead{(yr)\tablenotemark{2}} 
}
\startdata
J050629     & 154    & $-$0.150 & 0.688 & 54485 & 60273 & 15.8 \\
WBP~29      & 246    &    0.076 & 0.271 & 53483 & 60274 & 18.6 \\
MSX~LMC~736 & 690    &    1.375 & 1.057 & 54611 & 60385 & 15.8 
\enddata
\tablenotetext{1}{Modified Julian Date.}
\tablenotetext{2}{Time between IRS and MRS observations.}
\end{deluxetable*}

All of the photometry in Figure~\ref{f.cc} are from the Infrared Array
Camera on Spitzer \citep[IRAC;][]{faz04} and the Wide-field Infrared 
Survey Explorer \citep[WISE;][]{wri10}, supplemented with additional 
epochs obtained from the Near-Earth Object WISE Reactivated mission 
\citep[NEOWISE-R;][]{mai14}.  The IRAC data come from the SAGE survey 
of the LMC \citep[Surveying the Agents of a Galaxy's Evolution;][]{mei06} 
and its follow-up to study variability in the core of the LMC 
\citep[SAGE-Var;][]{rie15}.  In Figure~\ref{f.cc}, the WISE data at 3.4
and 4.6~\mum\ were shifted to the IRAC filters (3.6 and 4.5~\mum)
using the color-based corrections from \cite{slo16}.  The photometry
for all available epochs are averaged after color correction.

The base sample in Figure~\ref{f.cc} includes all SRVs and Mira
variables identified in the LMC by the OGLE-III survey \citep[Optical
Gravitational Lensing Experiment;][]{sos09}.  Figure~\ref{f.cc} 
appeared in its original form for the SMC \citep{slo15}.  While the 
details differ in the two galaxies, most likely due to their different 
metallicities, both show the same striking dichotomy, with the SRVs and
Miras clearly divided by [3.6]$-$[4.5] color.  The pulsational 
properties of the stars are linked to the quantity and chemistry of the
dust they produce \citep{kra19}.  The weakly pulsating SRVs produce small 
quantities of dust, much of it SiC, while the strong pulsations of the 
Miras result in higher dust-production rates dominated by amorphous carbon.

The full MRS sample follows the two sequences defined by the SRVs and 
Miras.  While the SRVs tend to be blue in most infrared colors, the 
[5.8]$-$[8] color stands out as an exception.  Because the stars are 
relatively dust-free, deep molecular bands can influence their color.
The culprit is likely to be C$_3$ at $\sim$5~\mum\ with a possible 
contribution from CO, which has a bandhead at 4.6~\mum\ \citep{slo15}.

\section{Spectra from the MRS on the JWST} 

\begin{figure}[!ht] 
\includegraphics[width=3.4in]{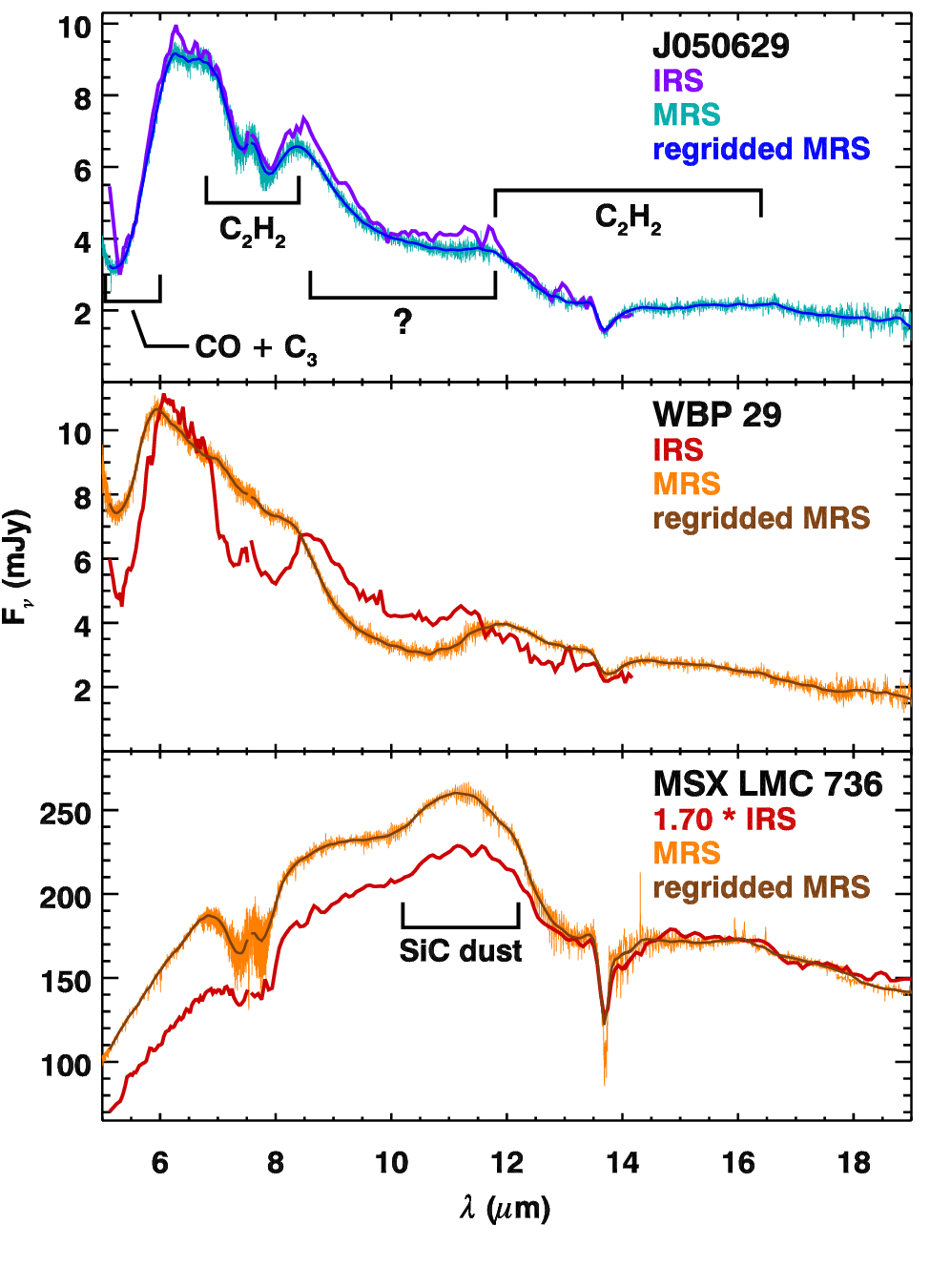}
\caption{MRS spectra of the three targets observed prior to 2024 June
compared to the spectra from the IRS.  The spectra from the MRS are also
plotted after downsampling them to the lower-resolution wavelength grid
for the IRS.  For MSX LMC 736, the IRS spectrum has been multiplied by
1.7 to align it to the MRS data at 17~\mum\ for easier comparison.
\label{f.spec}}
\end{figure}

Each MRS target was observed in all three grating settings to produce a
continuous spectrum from 5 to 28~\mum, although the quality of the data
deteriorates past $\sim$20~\mum, due to dropping sensitivity and, for the 
bluer stars in the sample, decreasing signal from the target itself.
The spectra in Figure~\ref{f.spec} were produced by the default JWST 
pipeline \citep{bus24}, which does not include a correction for the 
residual fringing.  The spectrophotometric calibration of the MRS agrees
with the IRS on Spitzer to $\sim$2\% or better \citep{law24}.

For each source, Figure~\ref{f.spec} shows a spectrum from the MRS at 
its full resolution, a spectrum of the same target obtained $\sim$16--19
years earlier with the IRS on Spitzer \citep[see][for details of the 
IRS observations]{slo16}, and the MRS data resampled to the IRS wavelength 
grid.  While some residual fringes remain in the MRS data, nearly all of 
the detailed structure in the full-resolution spectrum is the actual line 
structure within the molecular bands.  The apparent position of the 
``continuum'' in the downsampled version of the MRS spectrum demonstrates 
the challenge of determining the actual continuum in these complex spectra.

\vspace{0.2in}

\section{Light Curves and Pulsation Periods} 

For each target, we used the multi-epoch photometry from the SAGE surveys
at 3.6 and 4.5~\mum\ and WISE at 3.4 and 4.6~\mum\ to construct light
curves.  The IRAC data were shifted to the WISE filters using color
corrections algebraically determined from those provided by \cite{slo16}.

Table~\ref{t.sample} gives the pulsation periods for the three stars
investigated here.  For the two bluer sources, we used the periods 
provided by previous surveys.  The period of 154~d for J050629 comes from 
OGLE III \citep{sos09}, but OGLE III also suggested a period of 741~d with 
a slightly higher amplitude.  The MACHO survey \citep[the survey for 
Massive Compact Halo Objects;][]{fra05} fitted two periods with similar 
amplitudes:  267~d and 238~d.  None of these alternate periods fits the 
WISE and IRAC photometry for J050629 particularly well.  For WBP~29, we 
took the average from the MACHO and OGLE-III surveys (246 and 247~d, 
respectively).  

The light curves for J050629 and WBP~29 reveal a problem with our
conversions of the IRAC photometry at 3.6 and 4.5~\mum\ to the WISE
filters centered at 3.4 and 4.6~\mum.  The supposedly corrected IRAC
photometry is roughly 0.2~magnitudes brighter than the WISE photometry.
A comparison of contemporaneous data at MJD $\sim$ 55400 shows the 
issue most clearly.  The probable cause of the discrepancy is the molecular
band absorption affecting wavelengths outside the region of overlap in
the two filter sets.  We did not use the IRAC photometry when fitting
light curves for these two sources.

\begin{figure}[!ht] 
\includegraphics[width=3.4in]{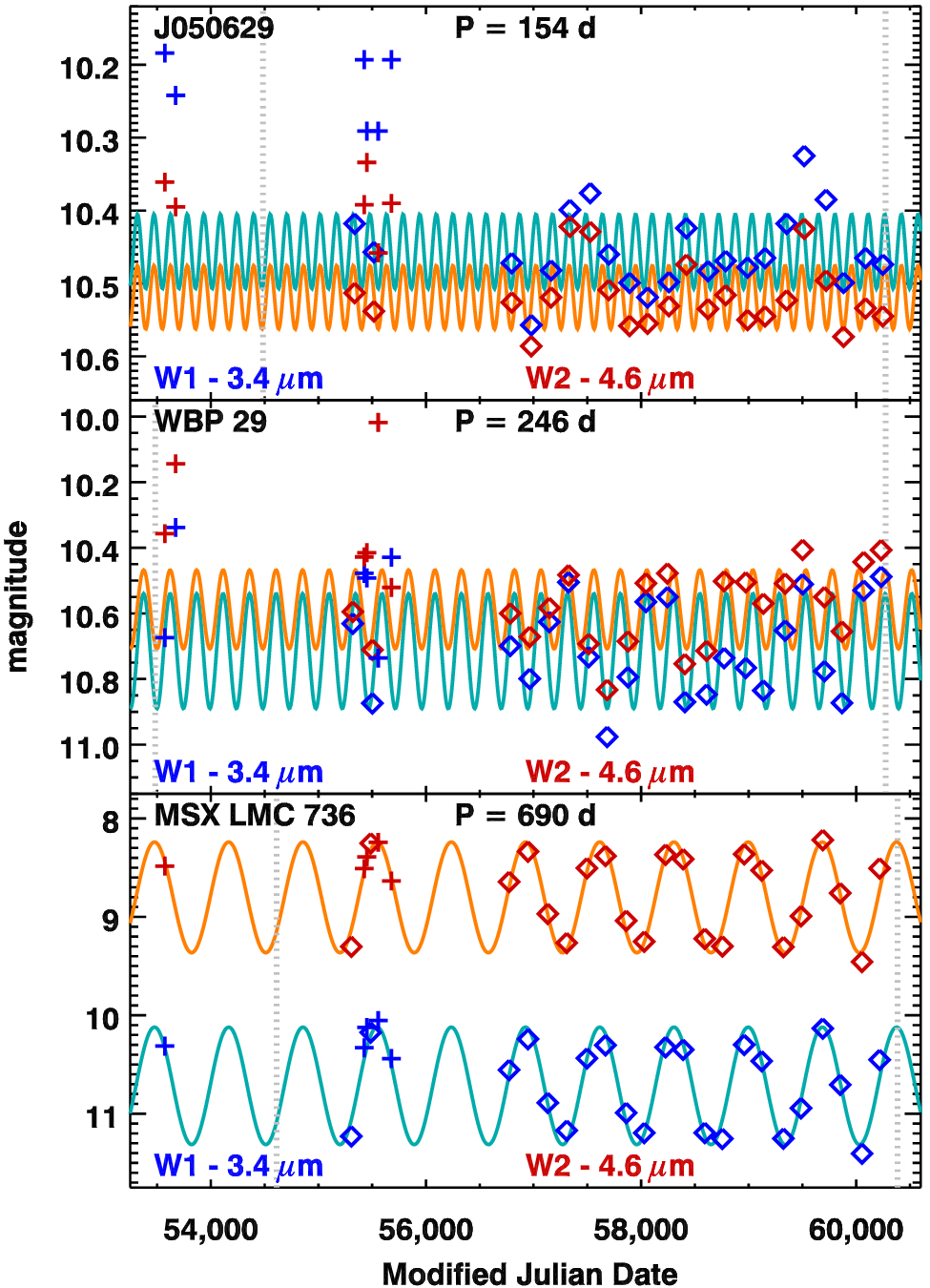}
\caption{Light curves for the three targets in the current sample.  The 
vertical dashed lines for each target mark the times they were observed 
by the IRS on Spitzer (left) and the MRS on the JWST (right).  WISE data are 
plotted as diamonds, and IRAC data (after conversion to the WISE filters) 
are plotted as plus signs.  The orange and light blue curves are fitted
to the red and blue data, respectively, as explained in the text.
Section \ref{sec5} discusses the reliability of the estimated phases 
for the MRS and IRS observing epochs for the three targets.
\label{f.lc}}
\end{figure}

For both J050629 and WBP~29, the adopted period was fitted to the WISE 
data with a sine function to determine the zero-phase epoch and 
amplitude for the two filters.  The zero-phase epoch was then averaged 
between the two filters to produce the light curves in Figure~\ref{f.lc}.
As explained in the next section, the estimated phases for the MRS and 
IRS epochs are not reliable for J050629 and are uncertain for WBP~29.

The OGLE-III survey is biased against the carbon stars most deeply 
embedded within their own dust shells, because the dust extinction makes 
them too faint to be detected in the optical.  The result can be seen in 
Figure~\ref{f.cc}, where the OGLE-based sample fades away at the reddest 
colors, even though plenty of targets were still observed by the IRS on 
Spitzer.  

Our reddest target, MSX LMC 736, is not in the OGLE-III sample, so we 
determined a pulsation period of 690 d from the multi-epoch WISE and 
(filter-corrected) IRAC data using the minimization algorithm described
by \cite{slo16}.  In brief, that algorithm iterates through periods,
fits an amplitude and zero-phase epoch to the photometry at each period, 
and returns the period with the smallest root-mean-squared residuals.
The result was consistent for both WISE filters and with 
and without the IRAC epochs (689.7 to 690.3~d).  Previous efforts using 
similar data and methods produced periods of 686~d \citep{slo16} and 
683 d \citep{gro18}.  The differences likely arise from fewer epochs 
available previously.  Using completely independent data from the VISTA 
Magellanic Cloud survey, \cite{gro20} found a period of 672~d.  The 
differences between these reported periods give some idea of the overall 
uncertainties.  Using a period of 690~d, a light curve was fitted to the 
available data as with the other two targets.  For MSX LMC 736, the 
color-corrected IRAC data were also used.

\section{Discussion}\label{sec5} 

\subsection{J050629} 

J050629 is the bluest target in [3.6]$-$[4.5] of the nine in the full 
MRS sample.  Of the three SRVs, it is the reddest in the [5.8]$-$[8]
color, and Figure~\ref{f.spec} shows why.  It has strong C$_3$ and CO
absorption at 5~\mum, and that is affecting the {{measured [5.8]$-$[8]}}.

The spectrum of J050629 has changed little between the IRS and MRS 
epochs.  Both epochs show an absorption band at 5~\mum.  Both spectra 
also show an acetylene (C$_2$H$_2$) band at 7.5~\mum\ with its 
characteristic ``W'' shape.  A second acetylene band appears at 
13.7~\mum, or more properly, the Q branch of a wider band that stretches 
from $\sim$12.5 to 16~\mum.  A broad, smooth absorption band from an 
unknown carrier is also present, centered at $\sim$10~\mum.  
A dust emission feature from SiC could be present at $\sim$11.5~\mum, but 
this apparent emission feature could just be the continuum between the two 
broad absorption bands to either side.  The most notable differences 
between the two spectra can be attributed to noise in the IRS data.  

While the light curve for J050629 suggests that the IRS observed it at 
minimum and the MRS observed it 15.8 years later at maximum, the short
pulsation period should limit confidence in that conclusion.  The
pulsation period is only 154~d, implying 37.4 pulsation cycles between
the spectral epochs.  Just a two-day shift in pulsation period would be 
sufficient to add half a pulsation cycle in 15.8 years, which would 
place both spectral observations at the same pulsation phase.  Given
that the star is a semi-regular variable, such a small shift seems
entirely possible.

\subsection{MSX LMC 736} 

MSX LMC 736 is the reddest target in [3.6]$-$[4.5] in the full MRS
sample due to the large quantity of amorphous carbon dust in its
circumstellar shell.  Its light curve shows strong pulsations, with
peak-to-peak amplitudes $>$ 1.0 magnitude in both filters.  Its 
spectrum shows both acetylene bands, centered at 7.5 and 13.7~\mum, 
as well as an unambiguous SiC dust emission feature at $\sim$11.5~\mum.

The MRS spectrum is 70\% brighter than the IRS spectrum, if one
compares the two at 17~\mum, but the MRS spectrum is even brighter
in the 5--12~\mum\ range, suggesting a higher fraction of warm dust.  
The long and steady pulsation period of MSX LMC 736 (Figure~\ref{f.lc}) 
results in a well-fitted light curve that shows clearly that the IRS 
data were obtained close to minimum and the MRS data close to maximum.  
The long pulsation period and the excellent agreement between the 
photometry and a sinusoidal model give us confidence in this conclusion.  
The changes in the spectrum are consistent with the star's pulsation 
cycle, with the MRS data obtained during a phase of higher luminosity of 
the central star, leading to higher atmospheric temperatures and warmer 
dust.

\subsection{WBP 29} 

Of the three stars in the sample observed before the Torino XIV AGB
meeting, WBP~29 is the enigma.  While the spectra from the IRS and MRS
have roughly the same brightness outside of the deeper molecular
bands, the bands have changed dramatically between the two epochs.
In the MRS spectrum, the C$_3$ and CO absorption at 5~\mum\ have been 
cut in half, while the 7.5~\mum\ acetylene band has nearly vanished.
In its place is a much stronger absorption band from an unknown carrier 
centered at 10~\mum\ (like the band in J050629, but stronger).  The low 
S/N at 14~\mum\ in the IRS data limits comparisons of the 13.7~\mum\ 
acetylene band.  The MRS data show a hint of what could be SiC dust 
emission at 11.5~\mum, but in all likelihood, that is just continuum.

WBP~29 is the bluest of the six Miras and closest to the boundary with 
the SRVs.  Its relatively blue color for a Mira indicates that it has
not yet produced a lot of dust, and that suggests that it may be in the
process of shifting from an SRV, as the amplitude of the fundamental
pulsation mode grows and begins to dominate the overtone modes.  As 
\cite{kra19} have shown, this shift is associated with an increase
in the dust-production rate for amorphous carbon.  In this context, 
the changes in the molecular chemistry could provide clues about how
the circumstellar envelope is changing as the star passes through this
transition to Mira.  This possibility was raised at the Torino XIV
AGB meeting.

If the phase information in the fitted light curve is reliable, then the 
IRS spectrum was obtained at minimum and the MRS spectrum at maximum.
That would mean that the spectral variations seen could be due primarily 
to the pulsation cycle.  However, the light curve fitted to the data for 
WBP~29 in Figure~\ref{f.lc} is not conclusive, and further analysis is 
needed for any certainty in the phases of the IRS and MRS observations.  
Whatever the cause of the differences in the spectra, the molecular 
chemistry has clearly changed.

\section{Conclusions} 

This short contribution describes the first three spectra obtained
out of nine in the total sample.  The analysis here focuses on the
overall shape of the molecular bands and dust features.  These new data 
demonstrate that temporal information is an important part of any study 
of carbon stars, due to their pulsations and dynamic behavior.  The 
molecular component in the new MRS data was the original objective of 
this project, and the data promise to reveal a great deal more about 
carbon stars as they evolve and ultimately eject their envelopes.


\begin{acknowledgements}

The authors thank the anonymous reviewers for helpful guidance on 
improving this paper.
K.~K., E.~M., R.~S., and G.~S. are supported by subawards from the 
Guest-Observer award JWST-GO-03010 provided by the Space Telescope
Science Institute.  R.~S.'s contribution was carried out at the Jet 
Propulsion Laboratory, California Institute of Technology, under a 
contract with NASA.  These results are based on observations with the 
NASA/ESA/CSA James Webb Space Telescope, which is operated at the Space 
Telescope Science Institute by the Association of Universities for 
Research in Astronomy, Incorporated, under NASA contract NAS5-03127. 
J.~C.\ acknowledges the support of the Natural Sciences and 
Engineering Research Council (NSERC) of Canada.
S.~H.\ acknowledges funding from the European Research Council (ERC) under 
the European Union’s Horizon 2020 research and innovation program (grant
agreement No.\ 883867, project EXWINGS) and the Swedish Research Council
(Vetenskapsradet, grant number 2019-04059).
M.~M.\ acknowledges support from the STFC Consolidated grant (ST/W000830/1).
This publication is based on work from COST Action CA21126{---}
Carbon molecular nanostructures in space (NanoSpace), supported by COST
(European Cooperation in Science and Technology).

\end{acknowledgements}


\end{document}